# Title: From Quantum Chemistry to Networks in Biology: A Graph Spectral Approach to Protein Structure Analyses

**Authors:** Vasundhara Gadiyaram[1], Smitha Vishveshwara[2] and Saraswathi Vishveshwara*[3]

1. IISc Mathematics Initiative (IMI), Indian Institute of Science,
   C V Raman Road Bengaluru, Karnataka 560012, India
2. Department of Physics, University of Illinois at Urbana-Champaign
   Urbana Illinois 61801-3080, USA
3. Molecular Biophysics Unit, Indian Institute of Science,
   C V Raman Road Bengaluru, Karnataka 560012, India

**Abstract:** In this perspective article, we present a multidisciplinary approach for characterizing protein structure networks. We first place our approach in its historical context and describe the manner in which it synthesizes concepts from quantum chemistry, biology of polymer conformations, matrix mathematics, and percolation theory. We then explicitly provide the method for constructing the protein structure network in terms of non-covalently interacting amino acid side chains and show how a mine of information can be obtained from the graph spectra of these networks. Employing suitable mathematical approaches, such as the use of a weighted, Laplacian matrix to generate the spectra, enables us to develop rigorous methods for network comparison and to identify crucial nodes responsible for the network integrity through a perturbation approach. Our scoring methods have several applications in structural biology that are elusive to conventional methods of analyses. Here, we discuss the instances of: (a) Protein structure comparison that include the details of side chain connectivity, (b) The contribution to node clustering as a function of bound ligand, explaining the global effect of local changes in phenomena such as allostery and (c) The identification of crucial amino acids for structural integrity, derived purely from the spectra of the graph. We demonstrate how our method enables us to obtain valuable information on key proteins involved in cellular functions and diseases such as GPCR and HIV protease, and discuss the biological implications. We then briefly describe how concepts from percolation theory further augment our analyses. In our concluding perspective for future developments, we suggest a further unifying approach to protein structure analyses and a judicious choice of questions to employ our methods for larger, more complex networks, such as metabolic and disease networks.

## I. Introduction

The scientific challenges of today, from the cellular level to neuroscience and physiology, to particle production in supercolliders, to disease propagation, to machine learning, have brought together the distinct disciplines of physics, chemistry, biology, mathematics and computational sciences. The exponential growth in the field of data science has not only provided eye opening knowledge, it has required identifying judicious choices of problems and developing interdisciplinary techniques to make progress. The problem of protein folding and structure analysis offers such a shining instance. Here, in the spirit of such multi-pronged approaches, we present a perspective of our interdisciplinary studies of Protein Structure Networks (PSNs).

Developed over decades, our comprehensive treatment of the PSN integrates principles from quantum chemistry to networks in a natural progression, building on developments from the past half a century. While the input in constructing the network formed of non-covalent interactions in protein structures is akin to quantum chemical methods to treat many electron systems, deriving the properties of these network structures hinges on powerful analyses based on matrix mathematics and percolation physics. A slew of graph theoretical measures such as hubs, communities, and paths of communication, serve to enhance our perception of protein structures. Further, matrix mathematics forms the basis for developing techniques for rigorous comparison of networks. In this perspectives article, we present a brief account of the development of such a program from spectral analysis of graphs (networks). Apart from demonstrating its utility in protein structure validation, we also provide insights through the identification of crucial players of network integrity and fundamental problems, such as allostery in biochemistry.

We begin here with a brief account of specific historical works that influenced the building of the graph spectral method for protein structure investigation. Central to our approach, graph theory has shed light on a host of key problems spanning multiple disciplines. Decades ago, the branch of mathematical chemistry (chemical graph theory) began to provide graph theory based heuristic methods to characterize the electronic structure and properties of small molecules[1,2], leading to the development and application of molecular descriptors in chemistry. During the 60's and 70's, quantum mechanical investigations of many electron systems of atoms and molecules (such as pioneered by one of our mentors, Sir John Pople, in the 1970's) gained momentum due to advances in molecular orbital theory and approximate heuristic schemes towards solving Schrodinger's equation[3]. In tandem, the rise of computers vastly enhanced calculation capabilities. The beauty of these computations was that the molecular orbitals were constructed from the atomic orbitals of electrons, thus enabling an interpretation of results such as, energy contribution, electron density, and their overlaps at atomic level within the molecular framework. Solutions emerged by treating the *n*- electron-density matrix as an eigenvalue problem. Today, ab-initio calculations of many electron systems has increased by several orders of magnitude, thanks to painstaking efforts by a large number of theoretical chemists from varying backgrounds and additionally, the exponential increase in computing power. The developments enabled scientists to gain a better understanding of the systems at hand and make predictions in various fields such as basic chemistry, material science, and biology. Finally, G. N. Ramachandran's approach to protein structures, based on simple non-covalent contacts, revolutionized the field of biopolymer conformation[4]. An amalgam of such concepts, techniques and emergent discoveries has led to the investigation of important problems in chemistry and biology through interdisciplinary approaches. Some of the most insightful approaches spanning quantum, soft, and biological matter have entailed seeking emergent organizing principles at the mesoscopic realm, intermediate between atomic and macroscopic scales[5], and they continue guiding us today.

Setting the stage for our investigations, the Ramachandran plot beautifully provides the underlying principle to comprehend the overall structure of a protein in terms of regular backbone secondary structures and the connecting loops. Today we try to seek answers to the question of whether there are any such guiding principles for obtaining a global view of the side chain connectivities in protein structures, as shown in Figure 1. This issue is important since the protein structure is encoded in its sequence[6] and the side chains are the key players in performing remarkable biological functions. Efforts are being made to identify general patterns in side chain conformations, similar to that of Ramachandran map [7]. Here we try to seek answers to the question of global connectivity of side chains in protein structures. Generally the side chain connectivity is treated at local levels. Based on our work for the past 2 to 3 decades, we are able to obtain some insights to the global connectivity of side chain in protein structures from network approaches. Further, we continue our efforts to unravel the role of side chain connections in encoding the stability to protein structures and their organization to perform their specific functions.

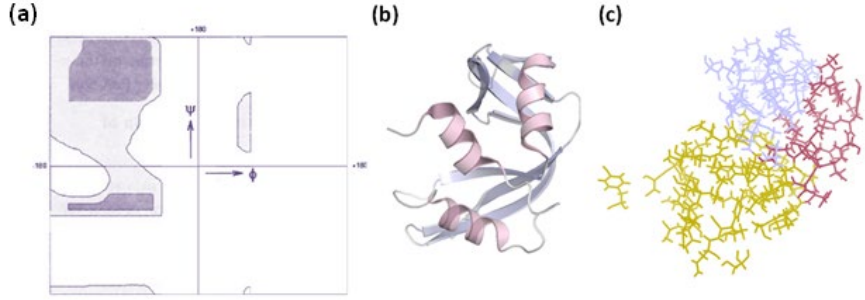

**Figure 1:** (a) G N Ramachandran map depicting the allowed and disallowed regions in the backbone torsion angles (ϕ, ψ) space; (b)Pymol representation of the backbone topology of a typical protein Ribonuclease-A(α-helices in pink and β-sheets in grey); (c) Three clusters of side chain connectivity in Ribonuclease-A, as obtained from the graph spectral analysis of the side chain network (described in section II)

The main focus of this paper is to offer a succinct review of our recent work on the protein structure network (PSN) employing graph spectral methods and to present a few key applications. In Section II, we first provide a brief account of earlier studies on the PSN formalism that are relevant to describe the latest development of graph spectral methods for protein structure comparison. All PSNs discussed here are constructed on the basis of non-covalent interactions (edges) between amino acid residues (nodes) in proteins. We then discuss the formulation of methods in terms of the transformations made on the non-covalent connectivity matrix, to obtain analytical solutions in the form of the spectra of the graph (network). Specific advances include the extraction of spectral features from graphs having weighted edges and employing normalized Laplacian matrices. Spectral features obtained from such matrices present an analytical way to compare any two networks, providing an excellent method of scoring and identifying the sources of differences between networks. Further, we make use of the scoring scheme to identify the key components responsible for the integrity of the network through perturbation studies. In Section III, we illustrate the power of this method through examples of influential proteins. We focus on one such case that is crucial for normal cellular functions, the beta-2 adrenergic receptor (a member of G-protein Coupled Receptor (GPCR) family), and one that is widely known for its involvement in the spread of deadly diseases, the HIV protease. In Section IV, we show how borrowing concepts developed in physics provides additional insights into PSN; we apply percolation theory analyses to the network structure to reveal that there are remarkable and universal aspects of its connectivity concerning the phase transition associated with a percolation threshold. In Section V, we conclude with a perspective for future developments, including the general application of the method to other disciplines, integration of further percolation concepts, and a judicious choice of questions to address larger, more complex networks, such as metabolic and disease networks.

**II. Graph Spectral Methods (GSM) for Analyses of Protein Structure Networks**

In this section, we provide (A) the relevant background to follow the review of recent developments on the adaptation of graph spectral methods (GSM) for protein structure networks (PSN) and (B) A succinct review of the methodological development.

**A. Background**

*Networks in the context of protein structure and function*

In the historical backdrop described in Section I, our modest contribution involves approaches towards understanding the structure-function relationship in proteins, the exploration of which began in late 90's. Akin to molecular orbital theory, where the molecules were treated as a linear combination of atomic orbitals, we treated protein structures as composites of amino acids (residues) in the polymer chain[8,9]. The folded structure of proteins was investigated at the non-covalent interaction level. Connections were made between interacting residues (considered as nodes) and edge-weights were assigned based on the extent of interaction, generating Protein Structure Network (PSN). The spectra of the connectivity matrices, which could either be in the form of 'Adjacency' or 'Laplacian', were analysed for properties of protein structures. The eigenvalues and the corresponding vectors of these matrices gave us a mine of information. For instance, the largest eigenvalues are associated with globally connected modes (analogous to the highest occupied molecular orbital (HOMO) state of electrons in conjugated molecules). The second lowest eigenvalue of Laplacian matrix represents the sub-clusters in the structures, as obtained from solutions to Kirchoff's matrix[10]. After establishing this basic methodology, biologically pertinent problems were investigated to extract relevant metrics for desired properties. For instance, identification of protein domains and their interfaces[11,12] or the center of a cluster, enables us to understand different modes of homodimeric association of lectins[13]. 'Kirchoff's matrix' was used for the first time by Bahar's group[14,15] to characterize the intrinsic dynamics of the protein structure through normal mode analysis, which led to the development of the Gaussian Network Model (GNM). The Protein structures have also been investigated for its network properties[16–20]. Several applications of analysis of side chain networks (from both conventional and spectral methods) relevant to protein structure-dynamics-function are presented in earlier review articles[21–23]

A distinct advantage of graph spectral analysis is that the spectra of networks capture maximum information with minimal loss. Recently we have introduced methodological developments in graph spectral methods to analyse protein structure networks in a systematic manner[24–26]. The specific features implemented in these studies are as follows: (a) Weighted edges in protein structures are used in constructing the Laplacian matrix for which an analytical solution is sought, (b) The normalized Laplacian matrices are utilized to develop Network Scoring Scheme (NSS) to compare different networks through the comparison of all eigenvectors of them and (c) Perturbation scores (NPS and EPS) are obtained by comparing the unperturbed network with the perturbed ones. Perturbation scores identify important players required for the network integrity. A brief description of these methods is provided below.

## B. Development of Methodology

In this section, we review recent advances [24–26] in PSN analyses using graph spectral features, incorporating the edge differences at the local level and the differences in modes of clustering at the global level. Some insights obtained on the network topology and its relationship with functions are also outlined. Further, the same principle utilized for network perturbation analysis is presented. It is to be noted that although the methods reviewed here focus on the PSN, they are general and applicable to any network having a defined set of entities (nodes) and their inter-connections (edges).

### *i) Definition of the Protein Structure Network (PSN)*

The details of constructing the edges and the corresponding matrices have been described in prior work[8,9,21–23] and relevant parts are presented here. Backbone networks, which have been successful in identifying gross features, such as overall topology and domain architecture, are characterized by identifying edges between non-covalently interacting $C_\alpha$ atoms within a distance of 6.5 Å. The crucial role of side chain atoms in determining the structure and function is efficiently captured by considering all atoms of the side chains. Hence, protein side chain networks are constructed with amino acid (residues) as nodes and non-covalent interactions among them as edges (at least one pair of their side chain atoms being within a distance of 4.5 Å). The interaction strength (edge weight) $I_{ij}$ is calculated using the equation

$$I_{i,j} = n_{i,j} / N_{ij} \qquad (1)$$

where $n_{i,j}$ is the number of atom pairs between residues *i* and *j*, within a distance cut-off of 4.5Å and $N_{ij}$ is normalization value which is the maximum possible number of contacts that the pair of residues (i,j) can make across a non-redundant database. The connectivity matrix thus obtained is known as the Adjacency matrix (A). The weighted matrix constructed here on the basis of geometric connections is simple, easy to generate, and amply used in earlier studies and in this article. However, the edge weights in the adjacency matrices can be generated by many ways such as those obtained from simulations or derived from statistical potentials.

For the sake of clarity, here we use the terminology PSN for networks both at the backbone level and at the side chain level with explicit inclusion of all side chain atoms. The terms PBN, and PScN refer specifically to the backbone network and the network constructed with all side chain atoms, respectively.

## ii) Network Comparison:

The development of the network comparison and perturbation scoring methods are described in detail in References (24 and 26) and a brief account of mathematical details is provided in the supplementary material (Section S1). Here we present the method of network scoring scheme (NSS) obtained from the spectra of weighted normalized Laplacian matrix and the interpretation of its components.

The NSS makes a rigorous comparison of two networks at both local and global levels. The unique feature of this method is to identify the best alignment of eigenvectors of the normalized Laplacian of the networks by comparing all eigenvectors with each other and quantifying the extent of alignment between them. Such a rigorous comparison ensures that the best alignments are captured. The components of NSS carry differences at various levels : (1) Correspondence Score (CRS) which represents the extent of eigenvector correspondence, capturing global level changes, (2) Eigenvalue Weighted Cosine Score (EWCS) which captures the extent of match between the best aligned eigenvectors, in other words quantifies the local node clustering, and (3) the Edge Difference Score (EDS), which represents the edge weight differences at the local level in the network. The key mathematical formulae of the components and NSS between two networks (network A and network B) are as follows.

$$CRS = 1 - \frac{6\sum(IndexEvec_A - IndexEvec_B)^2}{n(n^2-1)} \quad (2)$$

where $IndexEvec_A$ and $IndexEvec_B$ are the indices of the aligned eigenvectors of networks A and B, with maximum cosine values and $n$ is the size of the network (equal to the number of amino acids in the protein). We thus have

$$cosine(\theta_{ij}) = \frac{(Evec_i^A \cdot Evec_j^B)}{||Evec_i^A|| \, ||Evec_j^B||} \quad i,j \in N, \, 1 \leq i,j \leq n \quad , \quad (3)$$

$$EWCS = \frac{\sum(1-cosine)^2 |1-Eval_A| \, |1-Eval_B|}{\sum |1-Eval_A| \, |1-Eval_B|} \quad , \quad (4)$$

where $EVec_i^A$ is the $i^{th}$ eigenvector of network A and $EVec_j^B$ is the $j^{th}$ eigenvector of network B, which is aligned with $EVec_i^A$. The quantities $Eval_A$ and $Eval_B$ are the eigenvalues of $EVec_i^A$ and $EVec_j^B$ respectively.

$$EDS = ||M||_F / sqrt(\Sigma edge\text{-}weight_A \times \Sigma edge\text{-}weight_B) \quad (5)$$

where $A$ and $B$ are the adjacency matrices of network A and network B, $M$ is the difference of the two adjacency matrices and $||M||_F$ is the Frobenius Norm of the difference matrix $M$.

By combining the above components, NSS is written as the

Network similarity score (NSS) = $\sqrt{(1-CRS)^2 + (EWCS)^2 + (EDS)^2}$ (6)

It should be noted that a lower NSS reflects higher similarity between the reference and test networks. Furthermore, it should also be noted that the individual components that appear in the above equations not only contribute to the final score, but also carry information about differences at various levels (for example, local, global) of protein structures.

The scoring scheme and its implications are pictorially shown below for a representative protein structure in a summary (Figure 2). As illustrated, given two networks as input, the NSS captures the differences between them to a high degree of precision and also identifies local or global sources of differences, a challenge faced by network scientists. We present the validation of this method in Section III with illuminating examples from protein topology and function. The method described here can compare two networks of equal size. However, networks of different sizes can also be considered, which requires pre-processing to render them effectively of equal size by procedures such as introducing dummy nodes at appropriate positions.

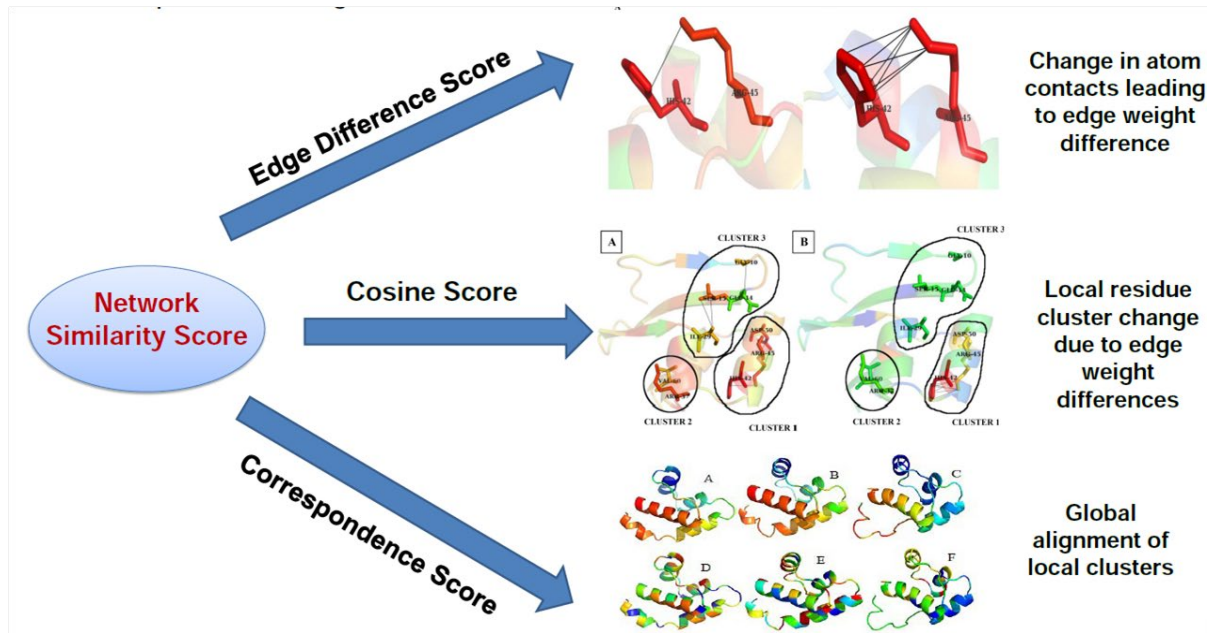

**Figure 2: Pictorial representation of the Network Scoring Scheme(NSS) and significance of the components**

*iii) Network Perturbation*:

The complex behaviour of any system, whether biological, social, or financial, emerges as a product of all the interactions between the components of the system as a single entity. Identifying crucial components which are more responsible for the integrity of the network is essential for managing or controlling these systems. Some of the standard methods for identifying such crucial players make use of 'degree' or 'centrality' measures. However, there are several limitations to these approaches (detailed in (26)), including the network not being considered to be a single entity. We have exploited the potential of the NSS for network comparison to identify the key players in maintaining the network architecture by considering all nodes to be part of the system[26]. Here the underlying principle is to characterize how properties change in the absence of a node or an edge in the network. The change due to the perturbation of a node or an edge in a network is estimated by quantifying the dissimilarity of the perturbed networks in comparison with the original one, using NSS, where the perturbations are created by systematic deletion of nodes/edges. The NSS thus obtained are used to derive perturbation scores[26].

The perturbation due to edge deletion and node deletion as evaluated from NSS, and are referred to as the edge perturbation score (EPS) and the node perturbation score (NPS), respectively. Each edge in the network is deleted one at a time and the resultant network is compared against the original one (undeleted network). Thereupon, the NSS is calculated. The resultant NSS values of all edges are normalized between 0 and 100. The edge perturbation score for a given edge 'e' takes the form

$$EPS_{(e)} = \{ NSS_{(e)} - \min(NSS)\} / \{ \max(NSS) - \min(NSS)\} \qquad (7)$$

where $NSS_{(e)}$ is the NSS value obtained from comparing the network, in which edge 'e' is deleted in the original network. The minimum and the maximum NSS values (min(NSS) and max (NSS)) are obtained considering all edges (1 to E) in the network.

Similarly, the node perturbation score for a node i ($NPS_{(i)}$) is evaluated by deleting the ith node and all its connections, and is given as:

$$NPS_{(i)} = \{ NSS_{(i)} - \min(NSS)\} / \{ \max(NSS) - \min(NSS)\} \qquad (8)$$

where min(NSS) and max(NSS) denote the minimum and maximum of NSS(i) of all nodes (1 to N) in the network. It should be noted that nodes/edges with higher perturbation scores (tending towards 100) are the ones whose deletion will cause more perturbation to the network. The significance of these scores is illustrated in the following section through the example of HIV protease.

## III. Illustrative examples highlighting the capabilities of the Graph Spectral Method

The graph spectral method (GSM) of network comparison (NSS) and the perturbation schemes (NPS,EPS) outlined above for the protein structure network (PSN) have great potential for a) large scale investigations related to characterizing the basic underlying principles in structural biology, b) employing as validation models, and c) generating extensive GSM data, which can further be used for structure prediction. In this section, we present a few examples to elucidate these points taken from recent works[24–26]. The first one demonstrates how new components to protein structure comparison can be incorporated. The second example elucidates the phenomenon of allostery, which is at times elusive to known methods of observation, for a working model of the ubiquitous family of proteins, G-protein coupled receptor (GPCR). As another example, to explore the effect of perturbation on the integrity of the side chain network of a protein, we have chosen the highly mutable protein HIV protease, which easily becomes drug resistant.

*i). Protein Structure Comparison*

Accurate structure validation of proteins is of extreme importance in a host of studies such as protein structure prediction, analysis of molecular dynamic simulation trajectories, and identification of subtle changes in very similar structures due to ligand binding, mutations, change in the environment, and more effects. Excellent validation scoring methods are available in the literature, including global distance test-total structure (GDT-TS), TM-score and root mean square deviations (RMSD). However, it is desirable to have a method that systematically respects all the side chain conformations (within the framework of backbone topology) and their interactions at the local as well as the global level. The NSS method meets this demand and has been shown to do an excellent job of rigorously comparing protein structure networks by considering all non-covalent interactions and their global repercussions. Here we present two examples of NSS applications to PSN comparison: a) Capability of the NSS to detect and score global features and b) Large scale validation of Critical Assessment of protein Structure Prediction (CASP) models and comparison of MD simulated structures to complement the existing methods with direct contributions from side-chain interactions (23).

**a*) NSS contribution highlighting the scoring of global effect*
As described in Section II.B.(*ii*), the uniqueness of the NSS lies in its ability to capture the differences between two PSNs at the global level by comparing the eigenvectors across the entire spectra (the terms CRS and EWCS in equation 6). The NSS results on CASP11 target TR821 (tetratricopeptide repeat protein) and two of the modeled structures (TS216_1 and TS396_3) showed that the all-atom–RMSD are almost identical in comparison with the native structure (Figure 3), while the side chain NSS score was significantly different. Surprisingly, the edge difference score (EDS) component was also very close for both the models, indicating negligible deviation in the non-covalent interactions. However, the major contribution to NSS scores came from the correspondence score (CRS) and the cosine score(EWCS) components (equation 6, Figure 2), clearly establishing the effect of the difference in local clustering and global alignment. This aspect was further probed through the Fiedler vectors (vector corresponding to the second lowest non-zero eigenvalue) of the spectra, to identify the specific regions contributing to the differences. Examination of the deviated nodes pointed towards the terminal helices in the structure. Minor local deviations, particularly at the hinge regions, that were not detected by any algorithm manifest themselves at a distant level, rendering a pair of side chain atoms of the two terminal helices connected. The clustering pattern was retained as in the native structure in the predicted model TS396_3 and deviated in the model TS216_1. Thus the spectral comparison not only provides the score difference, but also locates the regions causing the differences.

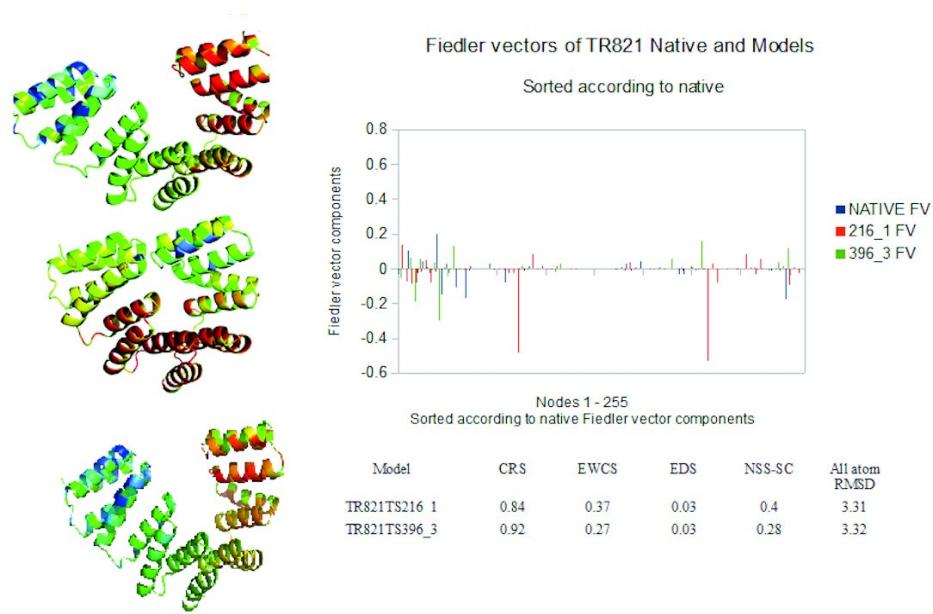

**Figure 3: (Left panel): Side chain clustering of TR821 native (top), TS216_1 (middle) and TS396_3 (bottom). In the model TS216_1, the two domains of the protein come close and atom-atom contacts are formed between the side chains from the two domains, which lead to side chain clusters (green-yellow), and a highly-connected side chain cluster (orange) which are otherwise absent in the native. The other model TS396_3 does not form such unusual clusters and is closer to the native compared to TS216_1. (Right panel): A plot of Fiedler vectors of the native and the two models, sorted according to native vector components. Model 216_1 contains more number of nodes (residues) getting deviated from native clustering(reference 25, figure adopted with permission to reproduce from (2019, John Wiley and Sons).**

*b) Large scale validation of CASP models and comparison of MD simulated structures*

A consolidated view on NSS performance is obtained by investigating datasets of different categories (crystal structures of lysozyme mutants, CASP11 refinement target models and MD generated trajectories(23) as given in Supplementary Material (Figure S1)). The observations are in correlation with RMSD profiles, as expected. For example, the range of backbone NSS are generally smaller than the side-chain NSS, indicating that there is more consistency at the fold level and diversity at the side chain connectivity level. Other expected features are that lysozyme mutants show low NSS values, consistent with minor deviations in the structures. The structures derived from high temperature simulations showed large deviations. Regarding CASP 11 models, some targets exhibited good performance (smaller NSS values) and the others showed relatively poor performance (higher NSS-backbone and NSS-side chains). Thus, NSS performance at the large scale is validated. Additionally it provides valuable information on the details of side chain interactions and their global effect when spectral metrics are analysed[25].

*ii) Allostery: Case study of Beta-2 Adrenergic receptor (β$_2$AR)*

Allostery, in simple terms, means action at distance. At the most fundamental level, conformational change induced by an appropriate ligand is generally accepted as the cause of allosteric communication. The extent of change could vary over a broad spectrum, ranging from drastic conformational change that can be observed by experiments to practically no apparent conformational change. Starting with works by Monod, Koshland, and co-workers[27,28], a large body of literature is dedicated to this subject [29–34] and interest still lives on. Here we focus on the contribution from the spectral method that provides valuable unique information on the changes in node clustering as a function of the nature of the bound ligand, which is elusive to many investigations. Although the structures provide the differences in edge weights between different liganded states of the protein, it is difficult to see the long distance effect of minor local changes. The unique advantage of the spectral method lies in accurate scoring in the case of comparison between large number of extremely similar networks and also identifying the regions of differences between the networks. The node clustering from the spectra of side-chain PSN can dramatically capture this global change. An example of Beta-2 Adrenergic receptor, a protein belonging to the GPCR family, is provided below to illustrate this point.

*a) Importance of G-Protein Coupled Receptors in Biology:*

Due to their implications in numerous biological phenomena, the transmembrane (TM) signalling molecules, G-Protein Coupled Receptors (GPCRs) are highly investigated and extensive reviews are available on the current status of GPCRs. A glimpse of the remarkable functions they are involved in is given here [35–41]. The proteins of this family possess extraordinary potential to respond to a diverse set of extracellular stimuli, such as light, ions, hormones, neurotransmitters and small molecule ligands and thereby mediate cellular signalling by interacting with heterotrimeric GTP-binding proteins (G-proteins). GPCRs control a variety of physiological processes that include sense of smell, taste, sight. They are also involved in immune response, behaviour, autonomous nervous system transmission, and homoeostasis modulation. At the heart of this signalling cascade lies the ligand-dependent activation of GPCRs, followed by G-protein coupling and nucleotide exchange that eventually culminates in regulation of downstream effector proteins.

The GPCRs function by means of ligand-driven activation followed by conformational changes that mediate interaction with GDP-bound G-protein heterotrimers ($G_{\alpha\beta\gamma}$). Nucleotide exchange and the subsequent dissociation of $G_{\beta\gamma}$ from $G_\alpha$ results in regulating the activities of cellular effectors, such as kinases, ion channels and other enzymes. Being one of the first GPCRs to be biophysically characterised and structurally determined by means of X-ray crystallography[42], the $\beta_2$-Adrenargic Receptor ($\beta_2$AR) serves as a classical system for understanding cell signalling. The $\beta_2$AR functions by binding to hormone and neurotransmitter adrenaline and inducing physiological responses, such as smooth muscle relaxation and bronchodilation via the agency of L-type Calcium channels[43].

b) *Graph spectral studies on Beta-2 Adrenergic receptor*

Here we focus on elucidating the spectral features[44] of ligand-induced conformational changes at side-chain levels on G-Protein+ agonist (POG) bound ($\beta_2$AR), an active conformation (PDB ID: 3SN6, hereafter referred to as $\beta_2$AR-$G_s$ )[42] and antagonist alprenolol-bound inactive state of $\beta_2$AR , (PDB ID: 3NYA, hereafter referred to as $\beta_2$AR-anta)[43] . The spectral decomposition of the Laplacian of weighted side-chain networks were carried out and the Fiedler vectors revealed the difference in clustering profiles of both the systems. The details are presented in Figures 4a and 4b.

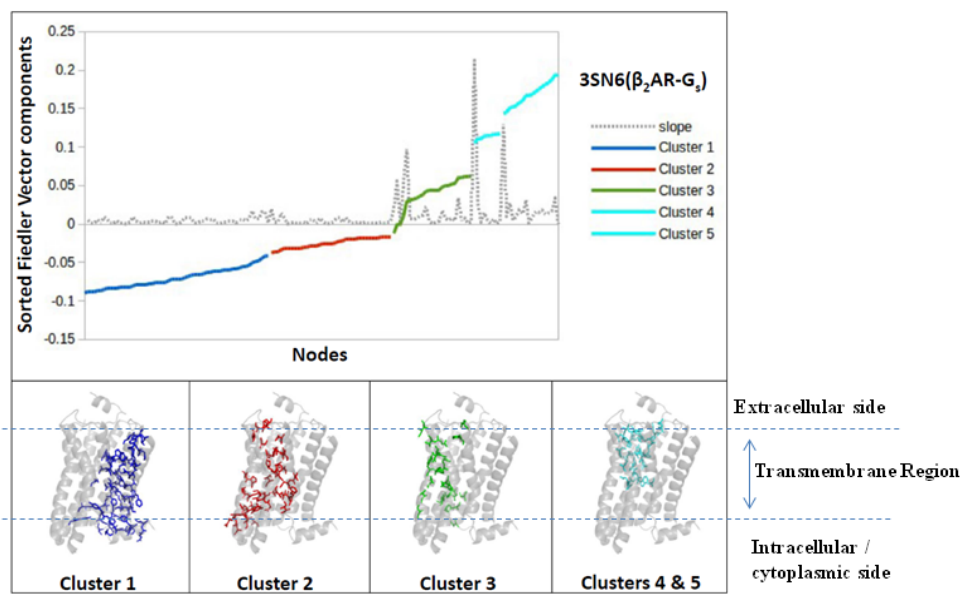

**Figure 4a:** Sorted Fiedler vectors(Fv) and clustering representations of $\beta_2$AR-$G_s$ (3SN6) Top: Plot of sorted Fiedler vectors in $\beta_2$AR-$G_s$ (3SN6); Bottom: Location of individual clusters mapped onto protein structure. In the figures 4a and 4b, the clusters are shown in different colours in the top panel, which is a typical plot that provides node clustering information. The magnitude of the nodes in Fv provides clustering information, where in principle, the nodes of the same or closely similar values belong to the same cluster. The slope of the graph helps in identifying clusters, with changes in the slope indicating cluster separation points. In the bottom panel, $\beta_2$AR is embedded in the transmembrane region, separating the extracellular and the intracellular sides of the membrane. The residues participating in each of the clusters are represented on the backbone.

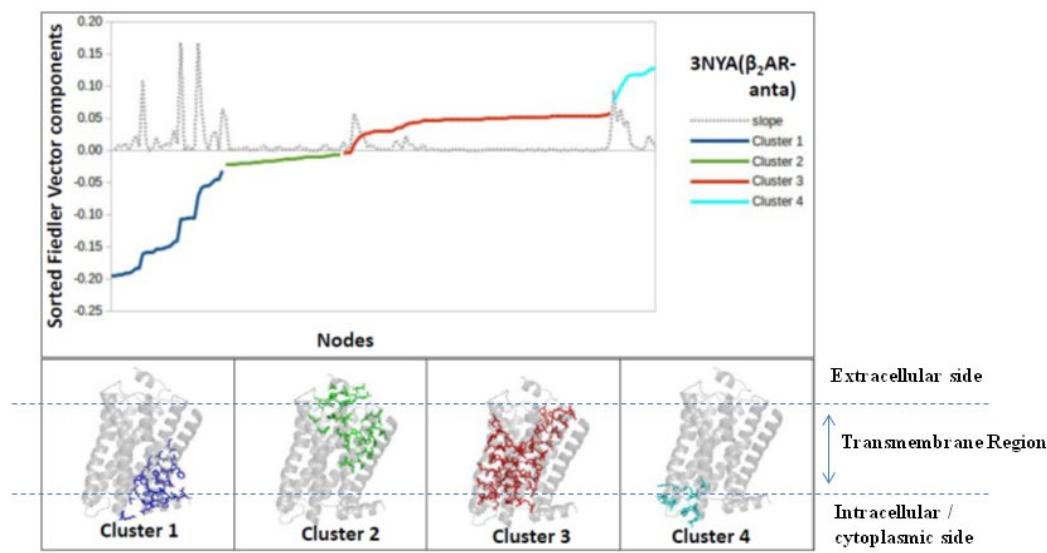

**Figure 4b: Sorted Fiedler vectors and clustering representations of β₂AR-anta (3NYA) Top: Plot of sorted Fiedler vectors in β₂AR-anta (3NYA); Bottom: Location of individual clusters mapped onto protein structure.**

The global effect of the disengagement of the TM helices in the agonist bound case i.e. β₂AR-G$_s$ (3SN6) and the proximity of helices in the antagonist bound case i.e. β₂AR-anta (3NYA) are interpreted in literature at the backbone helical orientation level. Here it is clearly elucidated at the side-chain interaction level, which is strikingly obvious from the clustering patterns adopted by these two systems. In the agonist bound case i.e. β₂AR-G$_s$ (3SN6), the residue clustering has happened in such a way that the clusters connect the extracellular and the intracellular regions of the membrane. This is reflected at the backbone level as the disengagement of helices in the agonist bound case, which is known to facilitate the binding of β₂AR to G$_s$ protein(Figure 4a). On the other hand, in antagonist-bound β₂AR-anta (Figure 4b, 3NYA), the clustering pattern is different, with residues of clusters 1 and 2 localized towards intracellular and extracellular sides of the membrane respectively. Such differences in node clustering can have dramatic impact in terms of keeping the receptor amenable for (or block) the communication across the membrane boundaries. In general, such changes in side chain residue clustering may not be apparent at the backbone level and often leads to interpretation as allostery with no conformational change. Thus, subtle changes in the conformations during allostery and protein-protein interactions are elegantly captured by difference in the edge weight and the manifestation of its effect at global levels.

### *iii) Identification of crucial players: Example of HIV Protease*

Real world networks are complex and the identification of crucial elements for network integrity is a highly non-trivial task. The complexity emerges as a product of all the interactions between the components of the system as a single entity and the contribution of each element towards network integrity may differ. The importance of each node in a network is assessed by various methods. Commonly used metrics in literature to identify the node contribution to network integrity include the degree of a node, path-associated metrics, such as betweenness[45] and closeness[46], and those related to stochastic processes, such as eigen centrality. The metrics employed naturally depend on the problem of interest. Most of the methods, however, lack the abstraction where the whole network is considered as a single global entity. Better solutions can be obtained by using methods which can consider the global connectivity of the networks at various levels. Since the spectra of a network capture global information, the graph spectral method was adopted to rank the nodes (Node Perturbation Score, NPS) and edges (Edge Perturbation Score, EPS) for their potential to maintain network integrity(24), as presented in Section II.A.(*iii*). Identifying such residues whose perturbation causes more changes in the network, especially in grouping of residues, can lend a helping hand in designing drugs, modeling synthetic proteins, controlling diseases and more[47,48].

The devastating consequence of the HIV infection is well known. To prevent the replication of HIV, a prevalent method involves inhibiting the enzyme HIV protease by using peptide-like drugs that bind in substrate binding site and suppress the conversion of HIV particles into their infectious form. But mutations that code for changes in conformational shape of HIV protease facilitate resistance of HIV to protease inhibitors. Details of these mutations are obtained from the Standford University HIV Drug Resistance Database[49] . These mutations are located primarily in the active site of the HIV protease, and also outside of the active site. The active site mutations directly change the interactions of the inhibitors and the non-active site mutations affect by other mechanisms like influencing dimer stability and conformational flexibility.

Here we make use of the node and edge perturbation scores (NPS and EPS) presented in Section II.B.(*iii*) to identify crucial nodes in the side chain network (PSN) of the protein HIV protease(PDB ID-1ODW). The nodes and edges having high perturbation scores (more influential in perturbing the network integrity) were evaluated and the top ranked nodes were identified(listed in Supplementary Material (Tables S1&S2)). The location of these nodes are shown in Figure 5b, on the dimer structure, in relation to structural topology, the residue conservation regions, and the positions of major drug resistant mutations[50].

The active site of the homodimeric HIV-1 protease includes the triad Asp25-Thr26-Gly27 from both the chains. Apart from this, several residues around the flap region (residues 49-52) and residues (81-90) that are spatially adjacent to the active site triad, also make contacts with the ligand as can be seen from Figures 5a and 5b. It has been pointed out that Thr26 is not found to interact with the ligand and hypothesized that Thr26 from chains A and B stabilize the conformational state of the active site through strong hydrogen-bonding forces[51]. Interestingly, the ligand binding residues such as Asp25 and several residues close to the flap region or the active site region are included in the top scoring node and edge perturbation list(NPS and EPS, Tables S1 and S2 in the Supplementary Material). Thr26 not only shows high node perturbation, but also shows a strong edge perturbation, where a strong edge is present between the Thr26 of the two chains of the (Figure 5a: Poseview Image of 478 in 3NUO [ taken from RCSB PDB])[52] dimeric protein.

Surprisingly, many of the nodes with high NPS or EPS are at the dimer interface or around the active site regions, which also host important (functional or conserved or drug resistant mutation) residues. This example indicates the utility of NPS and EPS in identifying critical components in the network, by quantifying the participation of nodes and edges. These results are mind boggling, considering the fact that only a tiny fraction of targeted drugs developed by drug industries make it to the clinical level, and beg for alternative approaches. The key message from this study is that a single chosen mutation can destabilize the structure network, leading to the loss of function; viral communities, very much including HIV, have learned this valuable lesson through evolution.

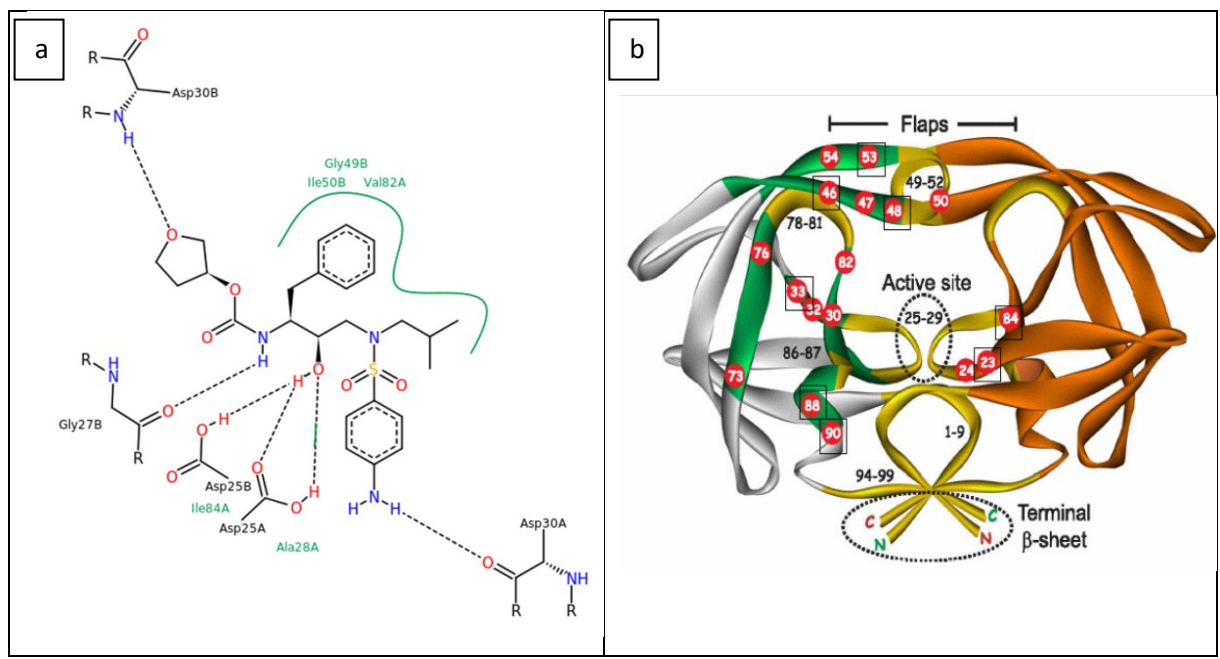

**Figure 5:** (a) Ligand (Amprenavir) interaction with the amino acid residues of the HIV-I protease, picture taken from: PDB-ID :3NUO (b) Ribbon representation of HIV protease, highly conserved regions(gold), naturally conserved regions where major drug resistant mutations are selected (green and orange). Numbered red circles indicate the positions of major drug resistant mutations, as defined in the Stanford database, Dimer interfaces are shown in dotted ovals. Black squares indicate residues with high NPS or involved in high EPS (reference 50, figure adopted with permission to reproduce from (2019, PNAS))

## IV. Application of Percolation Theory to Protein Structure Analyses

Having described the network and graph spectral approaches and demonstrated their effectiveness in two illustrative examples of key proteins, we briefly describe how integrating percolation theory, commonly used in physics, can be an attractive and powerful concept for further characterizing these PSNs. In the context of protein folding and function, reflecting perspectives from physics, strongly correlated building blocks can form phases of matter as a combined whole, such as the recently proposed[53,54] fascinating "elixir phase of chain molecules". Moreover, phase transitions are common at the molecular level and in biological systems, where many of the transitions pertaining to the same protein from one state to another decide the fate of the cell, including whether it be in a healthy or diseased condition. We find that the percolation perspective for exploring PSNs promises a better understanding of such critical physiological processes. The graph spectral description of the PSN established in Section II is highly amenable to percolation treatments, commonly used to study connected clusters in random graphs. We have investigated several features studied in percolation theory, compared with predictions for a large class of models, and offered perspectives on the biological implications of our results on the PSN and its formation[55,56].(Note that in this section, we refer to the side chain network as PScN and the backbone network as PBN, since both of them have been investigated). Specifically, the aspects we present here are as follows:

*a) Network properties and degree distribution*

Generally networks are characterized by their degree distribution, where the degree of a node refers to the number of links connected to it. For example, a large class of random models has degree distributions that peak around a specific value. In contrast, several real-world networks, such as the protein-protein interaction network, the World Wide Web, or the spread of diseases are considered to be scale-free in their degree distribution, a property that implies that certain nodes are highly connected[57,58]. PScNs (Protein side chain network) composed of non-covalent interactions differ from most known networks in terms of a) finite size in space corresponding to that of protein structure and b) each amino acid (node) constrained by the covalently connected polypeptide chain. Thus, assessing the network nature of PScN is not a straight forward procedure[17,55]. Additionally, PScN edges are weighted with a large number of weak edges and a few strong edges, and the degree distribution feature depends on the edge weight. Our study on a large number of protein structures of different sizes, showed complex behavior in the degree distribution (Supplementary Material Figure S3). PScNs mimic one of the simplest random network models, the Erdős–Rényi model[59], which has a degree distribution that peaks at a characteristic degree with a Poissonian form, in case of nodes with high edge weights. On the other hand, nodes including the weaker edge weights show an exponential behavior dominated with nodes of low degree with a small number of high degree nodes.

*b) Percolation Transition*

A hallmark of several random networks is the presence of a transition point at which a giant cluster percolates the system. In the Erdős-Rényi model[59] exact results are known; the percolation transition for a graph having a total of N nodes scales occur as a critical probability $p_c=1/N$. Below the transition, the largest connected cluster is small compared to N, but upon reaching the critical probability, sharply rises to a size of approximately $N^{2/3}$. To compare this behavior with our systems, we translate the interaction strength connecting one amino acid to another into a probability for two nodes of the PScN to be connected by a link. The PScN once again surprisingly exhibits Erdős–Rényi behavior; while it even closely conforms to quantitative results of the Erdős–Rényi model. The fact that a percolation transition exists in the PScN is significant in and of itself.

*c) Clique percolation*

In the global organization of many complex networks, structural subunits (also known as communities or modules) associated with highly interconnected parts, coexist. The overlapping of communities are shown to be significant and appear to be a universal feature of many real world networks such as protein-protein interaction networks[60]. An elegant method for capturing such presence of communities and their overlaps is through clique percolation[61].

A clique in a network consists of a cluster where each node is connected to every other node. If the number of nodes in a clique is 'k', a community is defined as the collection of adjacent k-cliques where each clique shares k-1 nodes with the adjacent clique. We identify the largest community as percolating clique. We investigated clique percolation features in the PScN and demonstrated (Figure 6, adopted from reference 56) it to have a significantly larger connectivity than random networks. It is in this measure that we discern palpable differences from the completely random, unphysical Erdős–Rényi model.

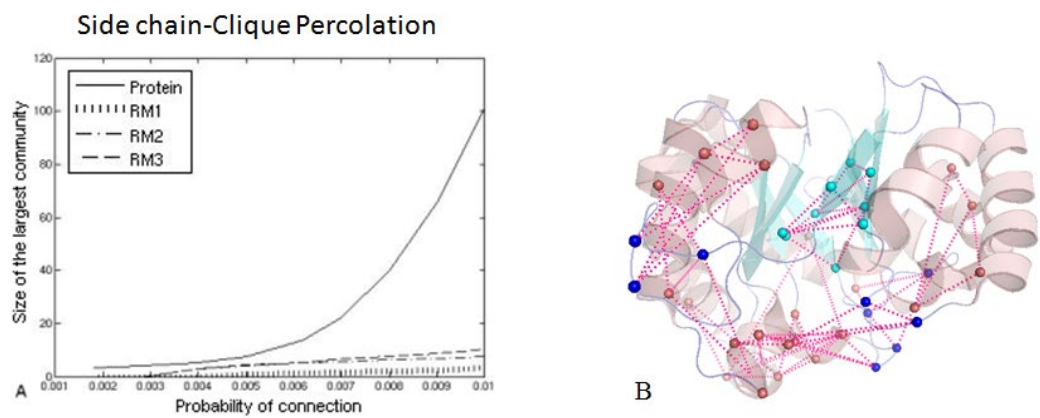

**Figure 6: (A)** Clique percolation profile of PScN. The inset is the key to the plots corresponding to different types of networks (of sizes ~400 nodes (RM1(Erdős–Rényi models), RM2 and RM3 (constrained random networks)). Number of nodes in the largest community (made from cliques of size 3) is plotted as a function of probability of connection. The side chain profile captures early stage of transition for PScN. On the other hand, none of the random models show any significant sized community within the probability of connections seen in real proteins (reference 56, reproduced with permission from (2019, Elsevier)); **(B)** Clique percolation in a typical globular protein (Triosephosphate Isomerase, a member of the TIM barrel fold: (PDBID=1LYX)), at Imin=3%, where the probability of connection is around the percolation transition point. [The helices, sheets and loops are shown in transparent red, cyan and blue respectively. The residues involved in the formation of percolating clique are shown as red, cyan and blue spheres. The non covalent connections among side chains of residues are shown as dotted lines in magenta colour)]; The communities become further connected when the probability of connection increases at the maximum possible side-chain connection (Imin=1%), permeating to cover almost the entire protein.

Furthermore, the clique percolation feature described seems to be universal to PScN (at least for globular proteins, which formed the dataset). This signature was also seen in different protein folds, some of which host very large number of diverse sequences.

Thus, the percolation studies on PSN show that the unique structure of a protein emerges from interplay between random and selected features. Specifically, the degree distribution and the bond percolation behaviour is distinctly reminiscent of random network models and the clique percolation specific to the side-chain interaction network bears signatures unique to proteins characterized by a larger degree of connectivity, than in random networks. These conclusions suggest that the randomness component seen in PScN may account for a large number of sequences with low homology being accommodated in the same protein fold. Our studies hint that while the element of randomness associated with the side-chain interactions constrained by the rigid framework of the protein backbone allow for functional flexibility and diversity, deviation from true randomness at the finer level reflects intrinsic uniqueness in structure and perhaps specificity in the functioning of biological proteins. Additionally, the randomness component seen in PScN may also account for a large number of sequences with low homology being accommodated in the same protein fold.

**V. Future Perspectives**:

Our interdisciplinary approach presented in this work on graph spectral characterization of protein networks offers several new directions. Here, we outline some possibilities with regards to direct applications, percolation studies, and complex biological networks.

1. **Direct applications of graph spectral method (GSM)**:

(a) The network comparison and the rearrangement of node clustering due to various perturbations in protein structures, captured by the graph spectral method described in this article, is a fairly straight forward application, if the protein structures are available at atomic resolution. For instance, a variety of functions in living cells are controlled by the G-protein coupled receptors (GPCR), as mentioned in Section III. An in-depth investigation of the available structures of these proteins in different liganded states through the graph spectral method would add new insights in terms of identifying side-chain interaction that are common to the class of GPCRs and unique to the specific receptors. Furthermore, the modes of node clustering at global level, occurring as a function of induced ligand type (such as, agonist, antagonist, and modulator) can be captured at the molecular level, thus enhancing our understanding of fundamental principles in biochemistry. Currently, some progress in this direction has been made in our laboratory. The method can also be applied to gain insights to a number of biochemical processes by providing well characterized structures of proteins and their interactions. Apart from crystal structures, comparison of simulation trajectories can offer the conformational landscape at the resolution of side chain connectivity. This multi-pronged approach can serve as an additional input for biophysical studies. Currently we are working towards developing a web server program based on the approach that will be made available to the community.

(b) The application of GSM is not limited to protein structures alone. One can consider any network of interest and convert the information into an adjacency matrix format to perform comparisons, characterize node clustering, or identify the most influential nodes that cause network perturbation at a global level. For instance, an application in structural biology would be to investigate proteins complexes with other macro molecules, such as nucleic acids as bipartite graphs, or to compare different structures of nucleic acids (DNA or RNA) for which the structure-function relationships are becoming available. Applications to disease networks can identify crucial proteins that can destabilize the normal functioning pathway.

**2. Integration of Graph Spectral Methods with Percolation Analyses**

An account of our attempts and the insights from the percolation studies on the PSN was presented in the previous Section (IV). In those studies, the clique/community percolations were characterized using a large number of experimentally observed (X-ray), folded-functional structures of proteins and the results were compared with different random models and decoy structures. Although these studies do not provide much insight into the mechanistic details of the process of protein folding (due to the lack of reliable structures of folding intermediates), they provided a tool to distinguish properly folded structures from decoys[62,63]. The range from valuable inputs for percolation parameters (specifically from side chain connectivities) to machine learning tools helped us rank the quality of modelled structures, which in turn enabled us to participate in the Community Assessment of Structure Prediction (CASP) experiments through the Wefold collaborative program[64]. Some of the successful structure prediction related to side chain interactions provided validation of our hypothesis that the signature of folded proteins lies in percolating clique-communities emerging from side chain interactions.

Further refinements are needed to improve the prediction score for capturing accurate side chain interactions of modelled or low resolution protein structures. Currently available protein structure prediction methods are performing an excellent job, as witnessed by a number of CASP experiments. Bringing the network perspective, particularly to fix side chain orientations in predicted structures will add a new tool to the analysis programs. In this context, the integration of recently developed graph spectral methods having metrics derived from percolation theory will provide new dimension for structure validation and serve as a robust method, which can be incorporated into structure prediction programs, reducing the structure space from the pool of predicted structures. Furthermore, at the conceptual level, the information obtained from currently available large datasets of protein structures can be investigated by GSM and integrated with percolation theory based metrics like communities, transition points to provide robust inputs to machine learning programs. By achieving this challenging task, new avenues will open up for more accurate prediction of a variety of functionally important structures, such as globular and membrane proteins, proteins folds with diverse sequences and diverse functions.

While the above studies are focused on intra-protein interactions of amino acids, they are subjected to the constraints of backbone architecture. Hence, the interpretation of network behaviour of PSNs become complicated as discussed earlier. A natural setting to dramatically experience the percolation effect is phase transition in molecules that are free from such constraints, examples being the interface interactions in multimers and aggregates of proteins. Nature offers many such instances, where monomer-multimer are in equillibrium, with multimers being the functional unit or giving rise to pathological conditions, as in amyloid proteins. Often a minor perturbation is enough to tip the equilibrium balance, leading to debilitating disorders, such as Parkinson's and Alzheimer diseases. Specifically, many intrinsically disordered proteins tend to aggregate and this phase transition triggers the disease states. Many of the structures and transition details are also available in literature. For example, Tau proteins are implicated in neurodegenerative disorders and they have also been characterized as two-liquid-phase states[65]. A better understanding of the process of phase transition and the structural organization in both states may emerge by investigating the percolating unit of monomer and the aggregate at molecular level, by specifically incorporating the details of side chain interactions. Here again, the computational advances mentioned in this article may play an important role in exploring such unchartered territories of molecular interactions.

3. **Exploration of complex biological networks**:

When a well characterized network is available, one can utilize a number of existing network algorithms or adopt graph spectral methods for extraction of network metrics to elucidate the property of the system. However biological networks are complex, since the concept of causality is not as precise as in the realm of physics, given that biology is heavily governed by evolutionary process. The highly quoted phrase from Max Delbruck[66] "Any living cell carries with it the experiences of a billion years of experimentation by its ancestors" is very relevant in this context. Evolutionary biologist Myer in his essay[67] has provided an excellent account of cause and effect in biology. He quotes that "Causality in biology is a far cry from causality in classical mechanics", however, he points out that "causality in biology is not in real conflict with the causality of classical mechanics" and it is the degree of unpredictability in physics (classical mechanics) and biology that are at the two ends of the spectrum. Furthermore he also adds that predictions of very high accuracy can be made with respect to most biochemical unit processes in organisms, such as metabolic pathways and most physicochemical phenomena on the molecular level.

Thus, the current challenge in dealing with biological networks at the organism level, is to establish the degree of uncertainty in different types of investigations, before they are integrated to a master network. In other words, serious thought has to be given to define a network and the type of questions to ask, to what extent the network can describe and explain phenomena with considerable precision, and to what extent the predictions from the analyses are reliable. This requires the functional and evolutionary knowledge of biological systems. Employment of concepts from physics, wherever is relevant, and the use of advanced mathematical/computer science techniques to enhance our understanding of a system from a more accurate and automated procedures are currently being explored. For example, the adaptation of percolation concepts (described above in the context of PSN) is also used in the context of disease module identification, to find out if the available data have sufficient coverage to map out modules associated with each disease [68]. Additionally, they have taken several steps to create an interactome (a network that integrates all known physical interactions within a cell) from several databases and made testable hypotheses like disease module hypothesis (network-based location of each disease module determines its pathobiological relationship to other diseases). Any new relevant technique can be explored in such a thought out large scale analysis. The spectra based network comparison scheme (NSS) and the perturbation scoring methods which are reviewed in this article can perhaps serve as independent validation methods, in the context of such exciting and challenging network approaches to diseases and medicine.

In summary, in this perspective we review our recent work on graph spectral method for protein structure networks and highlight its contribution in structural biology. Our presentation is embedded in the historical background of multidisciplinary approaches, spanning quantum chemistry, graph theory, percolation theory, and modern complex biological networks. We adopt matrix mathematics to extract maximum possible information from the spectra of networks, enabling rigorous characterization and comparison. We develop a network scoring scheme that incorporates the contributions from local edge differences and global alignment of local clusters. We apply the scheme to a perturbation method to assess the contribution of each element in the network for its integrity. As a specific instance, we obtain valuable insights on allostery in beta-2 adrenergic receptor from network spectra. We also evaluate perturbation scores on the structure of HIV protease to show crucial correlations between the network integrity score and mutational as well as biochemical effects. Our work presents a mesoscopic description of protein structure network. These investigations emphasize the fact that a holistic outlook of protein structure is required to understand fundamental principles and in practical applications from machine learning to drug design.

**Acknowledgements:**

We wish to dedicate this article to Sir John. A. Pople (1925-2004) and Professor David L. Beveridge, who are the postdoctoral advisor and the PhD mentor respectively, to Dr. Saraswathi Vishveshwara. We thank Dr. Vidhya Menon and Dr. Arinnia Anto for their great help in the preparation of the manuscript. Saraswathi Vishveshwara thanks National Academy of Sciences (NASI), Allahabad, India, for Platinum Jubilee Senior Scientist Fellowship. We acknowledge the Department of Biotechnology, Government of India for computing facility at the Molecular Biophysics Unit, Indian Institute of Science, Bengaluru, India

**Supplementary Material:**

**Section SI. Mathematical details used in Network Scoring Scheme (NSS) Method**

*Graph Spectra of Normalized Laplacian Matrix:*

The Laplacian matrix L(G) of a graph G is defined as

$$L(G) = DEG(G) - A(G) \quad (\text{or } L = D - A) \quad (1)$$

where DEG(G) is the degree matrix and A(G) is adjacency matrix of graph G.

The normalized Laplacian of a network correspnding to the adjacency **A** is given by

$$\mathbf{L^*} = \mathbf{D}^{-1/2} \mathbf{L} \mathbf{D}^{-1/2} \quad (2)$$

where D is the diagonal degree matrix defined as $D_{ii} = \deg(v_i)$, $v_i$ is any vertex in the adjacency matrix A and L is the Laplacian of A, where the matrix element $L_{ij}$ is given as

$$\mathcal{L}_{ij} = \begin{cases} -\dfrac{1}{\sqrt{d_i d_j}} & \text{if } i \text{ and } j \text{ are adjacent;} \\ 1 & \text{if } i = j \text{ and the vertex is not isolated;} \\ 0 & \text{otherwise.} \end{cases}$$

The eigen decomposition of the normalized Laplacian is carried out as follows

$$\mathbf{L}^* = \phi \Lambda \phi^T \quad (3)$$

where $\Lambda$ is the diagonal matrix containing normalized eigenvalues and $\phi$ contains the normalized eigenvectors. The eigenvalues of the normalized Laplacian of a symmetric matrix are real, non-negative and lie between 0 and 2. The vectors corresponding to these eigenvalues represent $n$ ($n$ is the size of the network) orthonormal vectors in an $n$-dimensional space into which the network is embedded. The set of these eigenvectors acts as a basis to the $n$-dimensional space and the vector components of each vector give the projections of the nodes on that dimensional vector. The set of eigenvalues and eigenvectors is considered as the spectrum of the graph, which is utilized further for network comparison.

**Figures and Tables**

(All reference numbers in this section correspond to the main manuscript)

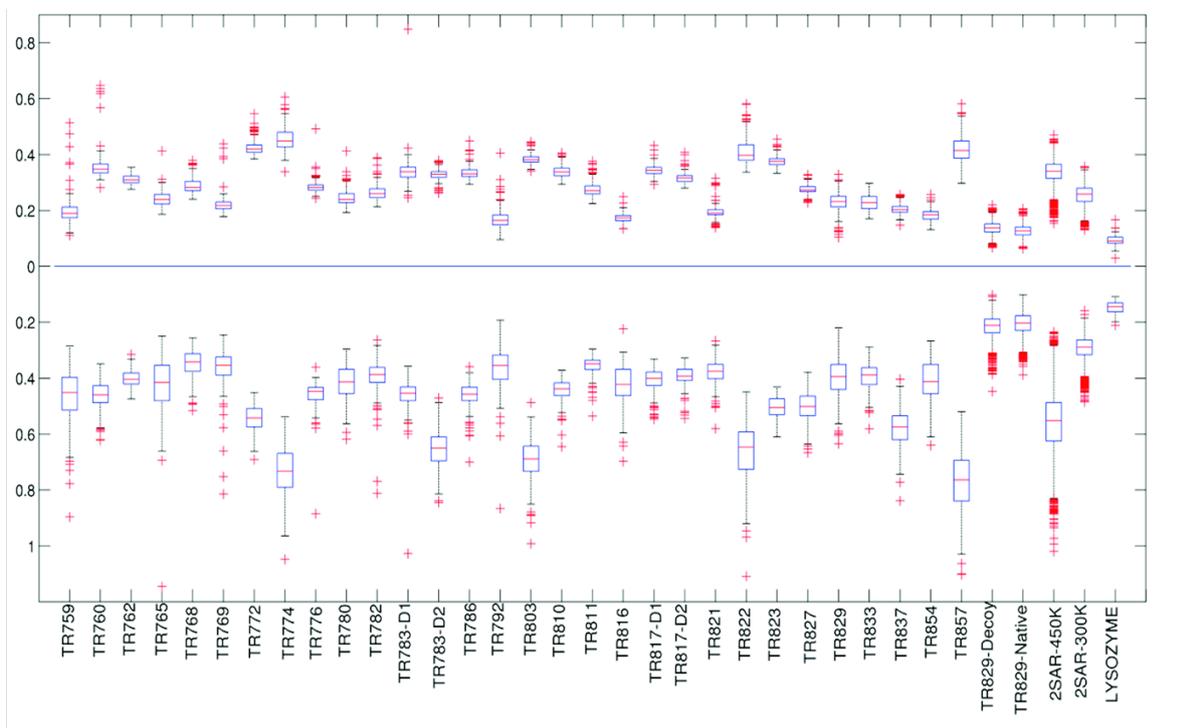

**Figure S1:** Box Plot of backbone and side-chain NSS scores. (Top panel): Backbone spectral score statistical distribution for all the test cases. (Bottom panel): Side-chain spectral score statistical distribution for the test cases (target models from CASP, simulation trajectories and lysozyme mutants). (Figure from Reference 25).

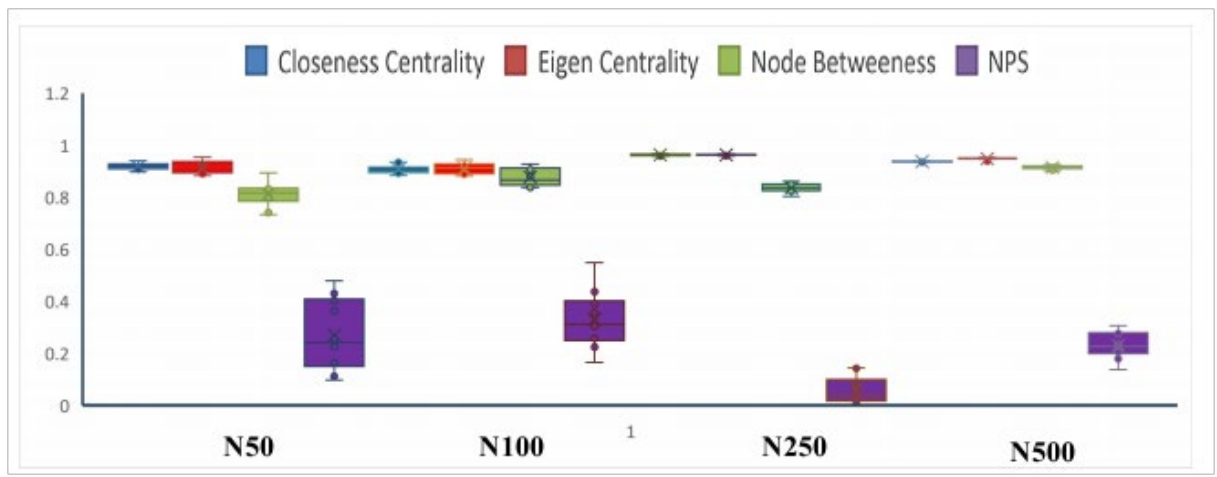

**Figure S2:** Correlation of closeness centrality, eigen centrality, node betweenness and NPS with degree of nodes.

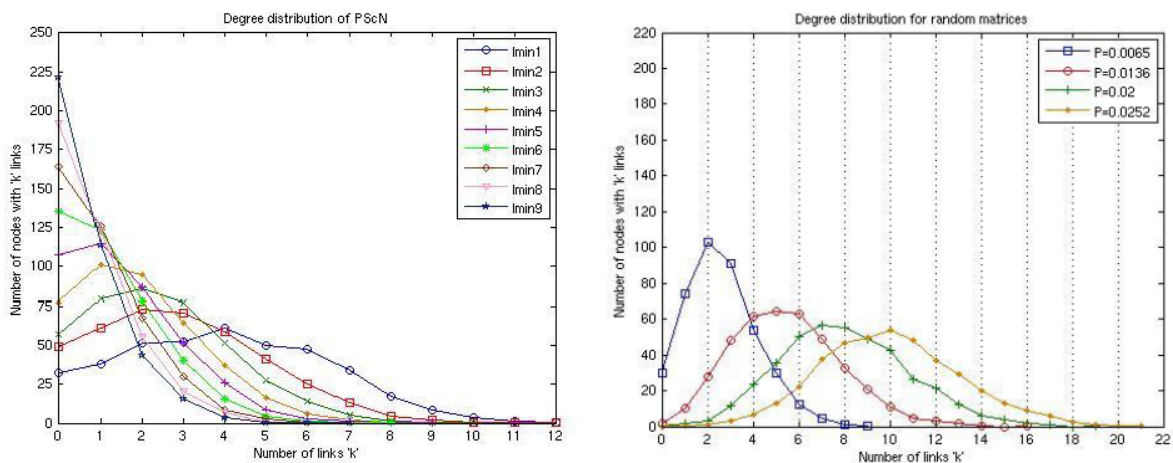

**Figure S3:** Degree distribution profile in PScNs [adopted from Reference 55]

Note: In the article we use the terminology PSN for both at the backbone level and at the side chain level with explicit inclusion of all side chain atoms. The terms PBN, and PScN refer specifically to the backbone network and the network constructed with all side chain atoms, respectively.

**Table S1: Residues with high NPS (top 14) in HIV protease and their significance**

| NPS | NODE | RESIDUE NUM | RESIDUE | CHAIN | SIGNIFICANCE |
|---|---|---|---|---|---|
| 100 | 25 | 25 | ASP | A | Conserved triad |
| 98.3 | 130 | 31 | THR | B |  |
| 83.2 | 46 | 46 | MET | A | Flap region |
| 81.7 | 127 | 28 | ALA | B | Active site residue |
| 80 | 124 | 25 | ASP | B | Conserved triad |
| 78.3 | 156 | 57 | ARG | B |  |

| 77.7 | 42 | 42 | TRP | A | |
|---|---|---|---|---|---|
| 72.3 | 84 | 84 | ILE | A | Substrate cleft |
| 72 | 132 | 33 | LEU | B | Major Non-cleft,non-flap drug resistant mutation |
| 71.8 | 23 | 23 | LEU | A | Substrate cleft |
| 71.7 | 147 | 48 | GLY | B | Substrate cleft |
| 71.2 | 56 | 56 | VAL | A | Flap region |
| 70.3 | 90 | 90 | LEU | A | Major Non-cleft,non-flap drug resistant mutation |
| 70 | 26 | 26 | THR | A | Conserved triad |

**Table S2: Edges with high EPS (top 14) in HIV protease and the significance of residues involved**

| EPS | NODE 1 | | | NODE 2 | | | EDGE WEIGHT | SIGNIFICANCE | |
|---|---|---|---|---|---|---|---|---|---|
| | RESNUM | RES | CHAIN | RESNUM | RES | CHAIN | | NODE1 | NODE2 |
| 100 | 46 | MET | A | 55 | LYS | A | 0.45 | Flap | Flap |
| 83.6 | 31 | THR | B | 89 | LEU | B | 0.7 | | Decreased PI susceptability |
| 76.0 | 29 | ASP | A | 87 | ARG | A | 0.3 | Active site | |
| 72.3 | 48 | GLY | B | 53 | PHE | B | 1.0 | Flap | Flap |
| 71.1 | 25 | ASP | B | 28 | ALA | B | 1.0 | Conserved triad | Active site |

| 67.1 | 14 | LYS | B | 65 | GLU | B | 0.25 | | |
| --- | --- | --- | --- | --- | --- | --- | --- | --- | --- |
| 65.8 | 28 | ALA | B | 86 | GLY | B | 1.0 | Active site | |
| 65.7 | 46 | MET | A | 53 | PHE | A | 0.2 | Flap | Flap |
| 65.5 | 22 | ALA | A | 85 | ILE | A | 0.5 | | |
| 64.0 | 44 | PRO | B | 55 | LYS | B | 0.2 | | Flap |
| 63.2 | 74 | THR | A | 88 | ASN | A | 0.1 | Decreased PI susceptability | Major Non-cleft,non-flap drug resistant mutation |
| 61.0 | 66 | ILE | B | 93 | ILE | B | 0.8 | | |
| 59.8 | 26 | THR | A | 26 | THR | B | 0.9 | Conserved triad | Conserved triad |
| 58.0 | 44 | PRO | A | 55 | LYS | A | 0.7 | | Flap |